\documentclass[twocolumn,floatfix,nofootinbib,10pt,twocolumn]{revtex4-1}
\usepackage{amsmath,amsfonts,amssymb,graphicx,times,bbm}
\usepackage{dsfont}
\usepackage{bbold}
\usepackage[usenames,dvipsnames]{xcolor} 
\usepackage{ulem} 
\usepackage{hyperref}

\DeclareMathOperator{\tr}{tr}

\hypersetup{colorlinks=true,
						linkcolor=MidnightBlue,        
						citecolor=MidnightBlue,        
						filecolor=magenta,      			 
						urlcolor=cyan                  
}

\begin{document}

\bibliographystyle{apsrev}
\newtheorem{theorem}{Theorem}
\newtheorem{corollary}{Corollary}
\newtheorem{definition}{Definition}
\newtheorem{proposition}{Proposition}
\newtheorem{lemma}{Lemma}
\newcommand{\proofend}{\hfill\fbox\\\medskip }
\newcommand{\proof}[1]{{\bf Proof.} #1 $\proofend$}
\newcommand{\nn}{{\mathbb{N}}}
\newcommand{\rr}{{\mathbb{R}}}
\newcommand{\cc}{{\mathbb{C}}}
\newcommand{\zz}{{\mathbb{Z}}}
\newcommand{\mbp}{\ensuremath{\spadesuit}}
\newcommand{\je}{\ensuremath{\heartsuit}}
\newcommand{\jd}{\ensuremath{\clubsuit}}
\newcommand{\id}{{\mathbb{1}}}
\renewcommand{\vec}[1]{\boldsymbol{#1}}
\newcommand{\me}{\mathrm{e}}
\newcommand{\mi}{\mathrm{i}}
\newcommand{\md}{\mathrm{d}}
\newcommand{\sg}{\text{sgn}}
\newcommand{\be}{\begin{equation}}
\newcommand{\ee}{\end{equation}}

\def\>{\rangle}
\def\<{\langle}
\def\({\left(}
\def\){\right)}

\newcommand{\ket}[1]{|#1\>}
\newcommand{\bra}[1]{\<#1|}
\newcommand{\braket}[2]{\<#1|#2\>}
\newcommand{\ketbra}[2]{|#1\>\!\<#2|}
\newcommand{\proj}[1]{|#1\>\!\<#1|}
\newcommand{\avg}[1]{\< #1 \>}

\renewcommand{\tensor}{\otimes}

\newcommand{\einfuegen}[1]{\textcolor{PineGreen}{#1}}
\newcommand{\streichen}[1]{\textcolor{red}{\sout{#1}}}
\newcommand{\todo}[1]{\textcolor{blue}{(ToDo: #1)}}
\newcommand{\transpose}[1]{{#1}^t}

\newcommand{\magenta}[1]{\textcolor{Magenta}{#1}}

\title{Simulating Bosonic Baths with Error Bars}

\author{M.P.\ Woods,$^{1,2,*}$ M.\ Cramer,$^{1}$ and M.B.\ Plenio$^{1,2}$ }

\affiliation{
$^1$Institut f\"{u}r Theoretische Physik, Universit\"{a}t Ulm, Germany\\
$^2$Quantum Optics and Laser Science, Blackett Laboratory, Imperial College London, United Kingdom
}

\begin{abstract}
We derive rigorous truncation-error bounds for the spin-boson model and its generalizations to
arbitrary quantum systems interacting with bosonic baths. For the numerical simulation of such
baths the truncation of both, the number of modes and the local Hilbert-space dimensions is
necessary. We derive super-exponential Lieb--Robinson-type bounds on the error when restricting
the bath to finitely-many modes and show how the error introduced by truncating the local Hilbert
spaces may be efficiently monitored numerically. In this way we give error bounds for approximating
the infinite system by a finite-dimensional one. As a consequence, numerical simulations such as
the time-evolving density with orthogonal polynomials algorithm (TEDOPA) now allow for the fully
certified treatment of the system-environment interaction.
\end{abstract}

\maketitle

\date{\today}
\let\thefootnote\relax\footnotetext{$^*$Now at Centre for Quantum Technologies, National University of Singapore and Department of Physics \& Astronomy, University College London, UK.}

\textit{Introduction} -- Ideal quantum systems may be considered closed, undergoing textbook unitary
evolution. In any realistic experimental setup however a quantum system is open, that is, it interacts
with an environment composed of those degrees of freedom that are not under the control of the experimenter.
Hence the numerical and analytical description of the dynamics of a quantum system in interaction with
its environment is of fundamental importance in quantum physics.
The precise nature and composition of the system-environment interaction is generally not known, but
for a wide range of systems encountered in physics, chemistry, and biology, it is common to model the
environment as a continuum of harmonic oscillators, which interact linearly with the system. This
results in the paradigmatic spin-boson model that captures many aspects of the system-environment
interaction \cite{LeggettCD+87}.
The spin-boson model is exactly solvable only in the rarest of special cases and one is therefore
compelled to employ a variety of approximations and numerical descriptions in order to obtain the
reduced dynamics of the quantum system in question. Notable examples include those cases in which
the environment possesses a correlation time that is much shorter than the system dynamics {\it and}
the system-environment interaction is weak. Under these assumptions it is then well-justified and
customary to resort to the so-called Markov approximation which permits the derivation of completely
positive and linear differential equations, the Lindblad equation, for the quantum system alone \cite{RivasH12}.

However, settings of considerable practical importance may violate either or both of these assumptions and
require a more sophisticated treatment. The recently emerging interest in quantum effects in biological systems
provides a case in point \cite{HuelgaP13}. For instance, in typical pigment-protein complexes the dynamical
time-scales of the vibrational environment can be comparable or even slower than the quantum mechanical
excitation energy transfer dynamics. Moreover, in the limit of slow bath dynamics, perturbative treatments
of the coupling between system and environment cannot be used even if the system-bath coupling is intrinsically
weak. Consequently, steps have been taken towards the development of non-perturbative
and non-Markovian approaches for the description of the quantum system-environment interaction (see
\cite{HuelgaP13,PachonB12} for overviews of recent developments). However, the majority of these approaches
have in common that they exploit approximations that are not well controlled in the sense that no
rigorous error bounds on the simulation results are available. Hence these methods are not certified.

\begin{figure}[t]
	\begin{center}
	\includegraphics[width=0.9\columnwidth]{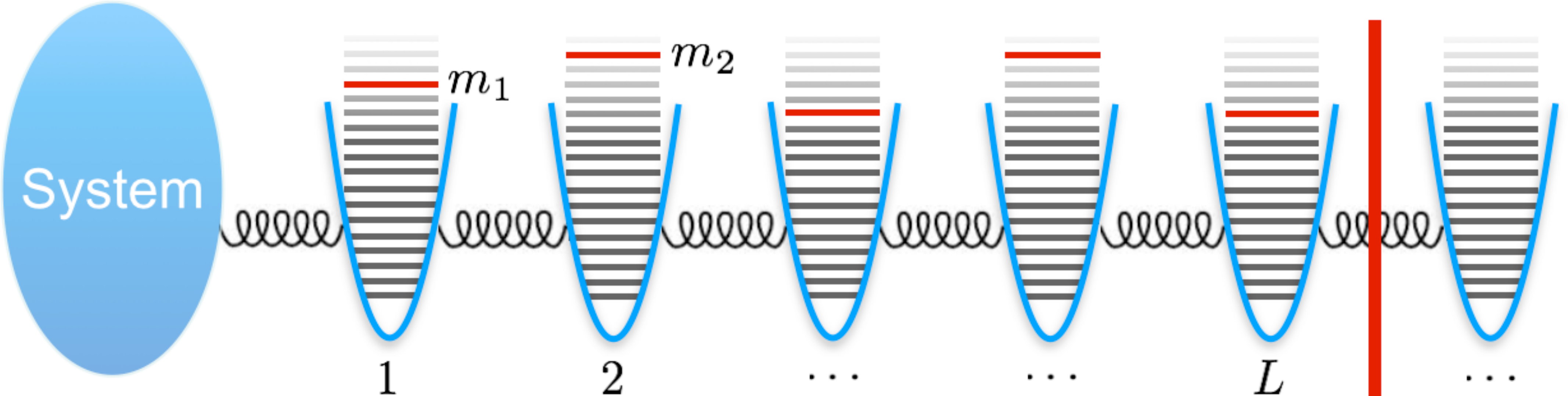}
	\end{center}
	\caption{A system coupled to a bosonic bath. Red lines indicate the truncations: The spatial truncation to a chain of finite length $L$ and the truncation of the local Hilbert space dimensions to $m_i$.}
	\label{cartoon}
\end{figure}

The time evolving density with orthogonal polynomials algorithm (TEDOPA) for the spin-boson model
presents a notable exception, as will be demonstrated in the present work. It makes
use of an exact transformation of the standard representation of the spin-boson model onto a spin
interacting with the first site of a semi-infinite nearest-neighbor coupled chain
\cite{burk84,prior10,alex10,mart11,misc14} which renders the system particularly amenable to time-adaptive
density matrix renormalisation group (t-DMRG) simulations. The structure of the resulting system is
such that excitations tend to propagate along the chain away from the system towards infinity leading
to irreversible system dynamics for long times. This approach has been used with success in the simulation
of a number of highly non-Markovian system-environment interactions \cite{prior10, prior13, alex13}.

The errors that accumulate in the t-DMRG simulation can be bounded rigorously. Nevertheless, the numerical
TEDOPA simulation employs two as yet uncertified assumptions: (i) the semi-infinite chain needs to be
truncated to a finite length and (ii) the local dimension associated with each harmonic oscillator of
 the chain the needs to be truncated to a finite dimensional Hilbert space, see Fig.~\ref{cartoon}.
The errors that are introduced in this manner are usually estimated by increasing both the chain length
and Hilbert space cut-off until the change in the result drops below a predefined threshold. However,
 in practice this somewhat inelegant approach can become highly challenging numerically,
and can lead to erroneous numerical predictions \cite{Volt09}. A more rigorous approach is therefore desirable.

Here we employ techniques that lead to Lieb--Robinson type bounds to achieve this goal by deriving bounds for the errors arising from approximations (i) and (ii). As the errors arising in each step of the t-DMRG integration can also be bounded we arrive at a method that possesses rigorous error bounds on the results that it delivers. This extends significantly existing recent results in the literature that apply to the finite dimensional setting of spin systems \cite{osbo06} and therefore allows the fully certified treatment of the system-environment interaction for both, harmonic oscillator as well as spin environments.

\textit{The system under consideration} -- We will consider the Hamiltonian of an arbitrary
system $\hat{H}_S$ coupled via $\hat{V}$ to a bosonic bath described by $\hat{H}_B$
so that the total Hamiltonian reads
\begin{equation}\label{tot ham}
    \hat{H} = \hat{H}_S + \hat{V} + \hat{H}_B.
\end{equation}
For simplicity and to directly connect to the TEDOPA approach \cite{prior10,alex10,prior13,alex13},
we assume that $\hat{H}_B$ describes a one-dimensional nearest-neighbour Hamiltonian (the higher
dimensional case with more general couplings will be published elsewhere \cite{future}) and takes
the form
\begin{equation}\label{the bath}
    \hat{H}_B=\frac{1}{2}\sum_{i,j=0}^\infty\left(\hat{x}_iX_{i,j}\hat{x}_j+\hat{p}_iP_{i,j}\hat{p}_j\right),
\end{equation}
where we assume that only nearest-neighbours are coupled, $X_{i,j}=P_{i,j}=0$ for $|i-j|>1$,
and we let  w.l.o.g. $X_{i,j}=X_{j,i}\in\rr$,  $P_{i,j}=P_{j,i}\in\rr$. We consider system-bath
couplings of the form $\hat{V}=\hat{h}\otimes\hat{x}_0$ (see the appendix for systems coupled
to several baths), where $\hat{h}$ acts on the system and we assume that it is bounded in operator
norm, $\|\hat{h}\|<\infty$. The system with Hamiltonian $\hat H_S$ has no restrictions, it can
correspond to any system---bosons, fermions, and/or spins, all in arbitrary dimensions.

\textit{Spatial truncation of the bath} -- For bounded system observables $\hat{O}$, $\|\hat{O}\|<\infty$,
We are interested in the quantity
\begin{equation}\label{delta def}
\Delta(t,L)=\bigl|\tr\bigl[\hat{O}\me^{-\mi\hat{H}t}\hat{\varrho}_0\me^{\mi\hat{H}t}\bigr]-\tr\bigl[\hat{O}\me^{-\mi\hat{H}_Lt}\hat{\varrho}_0\me^{\mi\hat{H}_Lt}\bigr]\bigr|,
\end{equation}
i.e., the error introduced when, instead of simulating the full Hamiltonian $\hat H,$ we simulate the time evolution of system observables $\hat{O}$ with the truncated bath Hamiltonian
\begin{equation}
\begin{split}
\label{eq:H_L}
\hat{H}^L_B&=\frac{1}{2}\sum_{i,j=0}^{L-1}\left(\hat{x}_iX_{i,j}\hat{x}_j+\hat{p}_iP_{i,j}\hat{p}_j\right)
\end{split}
\end{equation}
and corresponding total Hamiltonian $\hat{H}_L=\hat{H}_S+\hat{V}+\hat{H}^L_B$. Our first main result is the following.
\begin{theorem}\label{theorem 1} Let $\hat{H}$ and $\hat{H}_L$ be as above. Let $X,P>0$ or $X=P$ (see the appendix for a bound when neither of these conditions is satisfied). Let $c$ be such that $\|XP\|^{1/2}\le c$. Then
\begin{equation}\label{bound 1}
\frac{\Delta^2(t,L)}{4\|\hat{O}\|^2\|\hat{h}\|/c}\le C
\Bigl(\|\gamma_0\|^{1/2}+t\|\hat{h}\|
\Bigr)\frac{(ct)^{L+1}(\me^{ct}+1)}{(L+1)!},
\end{equation}
where $C=\|P_L\||X_{L-1,L}|/c^2+|P_{L-1,L}|/c$ and
\begin{equation}
\gamma_0=\left(\begin{array}{cc}
\gamma_{xx} & \gamma_{xp}\\
\gamma_{px} & \gamma_{pp}
\end{array}\right),\;\;\; [\gamma_{ab}]_{i,j}=\tr[\hat{a}_i\hat{b}_j\hat{\varrho}_0],
\end{equation}
collects the two-point bath correlations in the initial state. If $P\propto\id$, we may replace $L$ by $2L$ in Eq.~\eqref{bound 1}.
\end{theorem}
If the initial 2-point correlation functions (the matrix elements of $\gamma_0$) are unbounded, then one can still achieve bounds, see the appendix for details. The r.h.s. of Eq. (\ref{bound 1}) describes the Lieb--Robinson-type light cone \cite{nach10}. Outside the light cone, so for $\tau:=\me ct<L$,
one finds super-exponential decay in $L$: $(ct)^{L}\me^{ct}/L!\le \me^{ct-L|\ln(L/\tau)|}$.
This makes rigorous the physical intuition that for all finite times only a chain of \textit{finite} length is required to simulate the dynamics of local observables
to within a prescribed precision.
Our bound applies to \textit{any} system Hamiltonian, unbounded or otherwise, and depends only linearly on the operator norm of the system coupling $\| \hat h\|$.
The proof relies on Lieb--Robinson bounds for harmonic systems \cite{Marcus_bosonic_LR, Nachtergaele_bosonic_LR, ulma12} (see also Ref.~\cite{Alejandro}) and may be found in the appendix.
Before stating our second main result, we discuss the above bound in the light of the generalized spin-boson model.

\begin{figure*}
\begin{center}
\includegraphics[width=\textwidth]{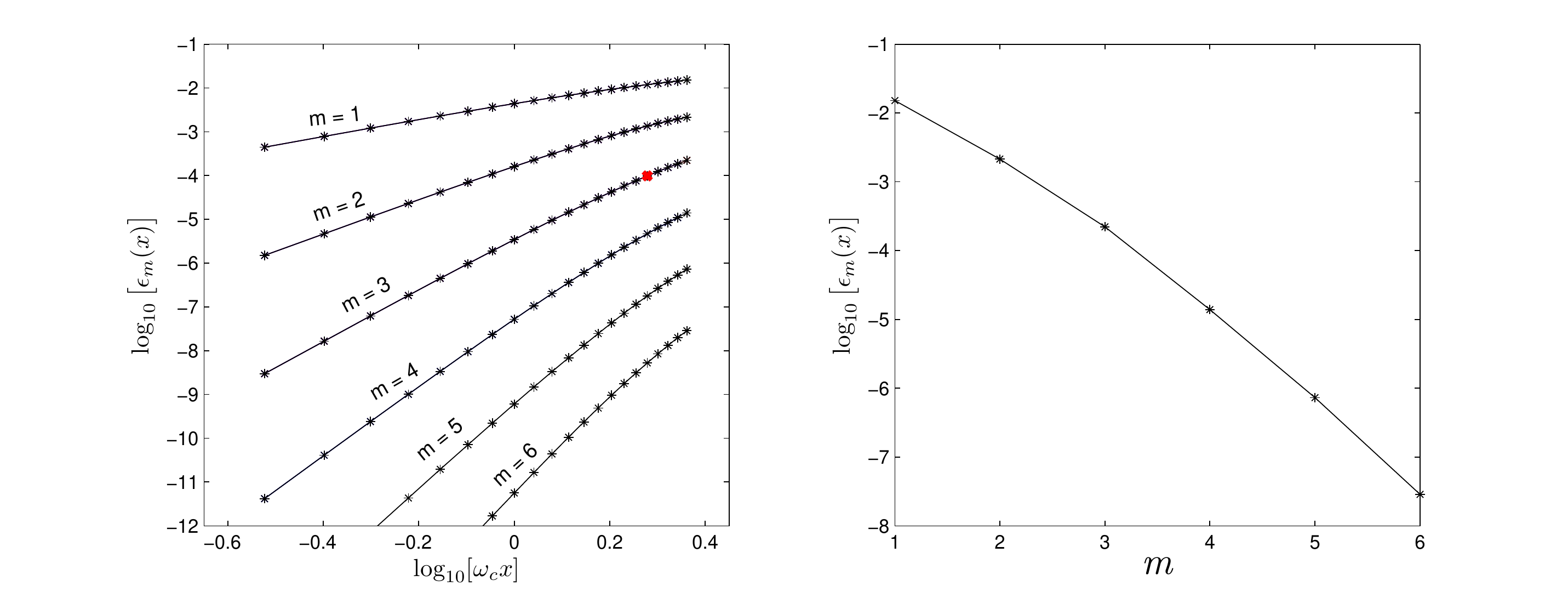}
\caption{Fock space truncation error (Eq. (\ref{ep x rho})) for the particle mapping and power-law spectral densities as in Eq.~\eqref{SB spec}
with $\Delta/\omega_{c}=1$, $\alpha=0.8$, $s=3$ for initial state $\hat{\varrho}_0=\hat{\varrho}_S^0\otimes\hat{\varrho}_B^0$, $\hat{\varrho}_S^0=|\!\!\uparrow\rangle\langle\uparrow\!\!|$ and $\hat{\varrho}_B^0$ the vacuum. We truncate each local Hilbert space at the same value $m_i=m$ and $L$ has the values $3$ to $6$, but are indistinguishable (e.g. the difference between the $L=6$ and $L=3$ curve at the point denoted by a red square is $4.95 \times 10^{-6}$). Lines are guides to the eye. The log-log plot on the left suggest algebraic increase in time and the plot on the right suggests better than exponential decrease with $m$.}
	\label{fig}
	\end{center}
\end{figure*}

\textit{Generalised spin-boson model} -- In this section we will investigate Hamiltonians of the form
\begin{equation}\label{the bathtub}
\hat H=\hat H_S+\int \md\vec k\, g(\vec k)a^\dagger_{\vec k} a_{\vec k} + \hat A_S\int \md\vec k\, h(\vec k) (a_{\vec k}^\dagger+a_{\vec k}).
\end{equation}
This describes a quantum system with Hamiltonian $\hat H_S$ interaction with a bath of bosons; it is described in more detail in terms of second quantised operators in \cite{dere99}. This model 
has received renewed interest in recent years due to its importance in the theoretical study of quantum effects in biology (see \cite{HuelgaP13} for a review). An important quantity that describes the bath and its coupling to the system is the spectral density, which, for invertible $g$, is defined as
\begin{equation}\label{SD_eq}
J(\omega)=\pi h^2\left(g^{-1}(\omega)\right)\ \left|\frac{dg^{-1}(\omega)}{d\omega}\right|,
\end{equation}
with $g^{-1}$ the inverse of $g$. The smallest closed interval containing the support of $g^{-1}$ is denoted $[\omega_{min},\omega_{max}]$. The case $\omega_{min}=0$ is called massless where as $\omega_{min}>0$ is known as massive.

Building on the work of \cite{burk84, mart11, alex10}, it was shown using the theory of orthogonal polynomials in \cite{misc14} that Eq.~(\ref{the bathtub}) can be written in the form of Eqs.~(\ref{the bath},\ref{tot ham})
and that there are two ways to do this. Both choices
\begin{equation}\label{particle rep}
\hat{h}=\mu_0\hat{A}_S,\;\;\; X=P,
\end{equation}
and
\begin{equation}\label{phonon rep}
\hat{h}=\mu_1 \hat{A}_S,\;\;\; P=\omega_{max}\id,
\end{equation}
with appropriate $X$ (given in terms of the spectral density in the appendix) are equivalent to Eq.~(\ref{the bathtub}). Here,
\begin{equation}\label{eq: mu 1 and mu 2}
\mu_0^2=\frac{2}{\pi}\int \md \omega\,J(\omega),\;\;\; \mu_1^2=\frac{1}{\pi\omega_{max}}\int \md \omega\,J(\sqrt{\omega})
\end{equation}
and one finds $\|X\|=\|P\|=\omega_{max}$ for both cases and $X>0$ iff $\omega_{min}>0$. Due to the
form of their elementary excitations, the mappings leading to couplings as in Eqs.~(\ref{particle rep}) and (\ref{phonon rep}) were named
\textit{particle mapping} and \textit{phonon mapping}, respectively, and we will adopt this denomination
here. Crucially, in both cases, $X$ couples nearest-neighbours only such that the bound in Eq.~(\ref{bound 1})
is readily applicable to the particle and the massive phonon case, setting $c=\omega_{max}$ for both
(similar results hold for the massless case, see appendix for full details). For the particle mapping,
we find $C\le 2$ and for the phonon mapping $C\le 1$ such that, up to the constants $\mu_{0/1}$, we
obtain the same behaviour of the bound in both cases but replacing $L$ by $2L$ in the massive phonon
case. Hence, for the phonon mapping with a chain of only half the length, one has approximately the
same chain truncation error  as for the particle mapping.

If the maximum frequency of the bath $\omega_{max}=\infty,$ the chain coefficients are unbounded
\cite{misc14} and our bounds diverge. This divergence is not surprising in light of the observation
that certain one-dimensional infinite harmonic lattice models with nearest neighbour interactions and
unbounded coefficients have been proven \textit{not} to have a light cone bound \cite{eise09}. It is
noteworthy, that similar results can be derived for the case of a fermionic bath, since the chain
mapping is still valid and Lieb--Robinson bounds for fermions are well-known \cite{LR4}.

\textit{Truncating local Hilbert spaces}\label{Truncating local Hilbert space} -- We now consider the error introduced when the local Hilbert space dimensions of the harmonic oscillators making up the bath are truncated. To this end, we define the projector
\begin{equation}
\id_{\vec{m}}=\id_{m_0}\otimes\cdots\otimes\id_{m_{L-1}},\;\;\;\id_{m}=\sum_{n=0}^{m}|n\rangle\langle n|,
\end{equation}
where $\id_{m_i}$ acts on the $i$'th site of the bath and truncates the local Hilbert space according to $\id_{m}$.
For bounded observables acting on the system $\hat{O}$, $\|\hat{O}\|<\infty$, we consider
\begin{equation}
\Delta_{\vec{m}}(t)=
\bigl|\tr[\hat{O}\me^{-\mi t\hat{H}}\hat{\varrho}_0\me^{\mi t\hat{H}}]-\tr[\hat{O}\me^{-\mi t\hat{H}_{\vec{m}}}\hat{\varrho}_0\me^{\mi t\hat{H}_{\vec{m}}}]\bigr|,
\end{equation}
i.e., the error
introduced by evolving the system according to
\begin{equation}
\hat{H}_{\vec{m}}=\id_{\vec{m}}\hat{H}\id_{\vec{m}}
\end{equation}
instead of $\hat{H}$. Here, $\hat{H}$ is as in Eq.\ (\ref{eq:H_L}) and we omit the index $L$ for notational clarity.
The truncated Hamiltonian reads
$\hat{H}_{\vec{m}}=\hat{H}_S+\hat{H}_B^{\vec{m}}+\hat{h}\otimes\id_{\vec{m}}\hat{x}_0\id_{\vec{m}}$, where
 \begin{equation}
 \hat{H}_B^{\vec{m}}=
 \frac{1}{2}\sum_{i,j=0}^{L-1}\bigl[X_{i,j}\id_{\vec{m}}\hat{x}_i\hat{x}_j\id_{\vec{m}}+P_{i,j}\id_{\vec{m}}\hat{p}_i\hat{p}_j\id_{\vec{m}}\bigr].
 \end{equation}
In the appendix we show that
\begin{equation}\label{delta m}
\begin{split}
\frac{\Delta^2_{\vec{m}}(t)}{4\|\hat{O}\|^2}\le
\tr\bigl[(\id-\id_{\vec{m}})\hat{\varrho}_0\bigr]+2
\int_0^t\md x\,\sqrt{\epsilon_{\vec{m}}(x)},
\end{split}
\end{equation}
where
\begin{equation}\label{ep x rho}
\epsilon_{\vec{m}}(x)=
\tr\bigl[\hat{h}^2
\me^{-\mi x\hat{H}_B^{\vec{m}}}\hat{X}^2(x)
\me^{\mi x\hat{H}_B^{\vec{m}}}\hat{\varrho}_{\vec{m}}(x)\bigr],
\end{equation}
with
\begin{equation}
\begin{split}
\hat{X}(x)&=\id_{\vec{m}}\me^{\mi x\hat{H}_B}
\hat{x}_0\me^{-\mi x\hat{H}_B}\id_{\vec{m}}
-
\me^{\mi x\hat{H}_B^{\vec{m}}}\hat{x}_0\me^{-\mi x\hat{H}_B^{\vec{m}}},\\
\hat{\varrho}_{\vec{m}}(x)&=\me^{-\mi x\hat{H}_{\vec{m}}}\hat{\varrho}_0\me^{\mi x\hat{H}_{\vec{m}}}.
\end{split}
\end{equation}
Crucially, under the assumption that the system Hilbert space is finite dimensional, this error may be computed numerically as it involves only observables acting on the truncated Hilbert space ($\me^{\mi x\hat{H}_B}\hat{x}_0\me^{-\mi x\hat{H}_B}$ is a linear combination of the $\hat x_i$ and $\hat p_i$)
and which are of a form amenable to t-DMRG simulations (see the appendix for details). For all finite times, $\lim_{ \{m_i\} \rightarrow \infty} \Delta_{\vec m}=0$ and we study its behaviour in $\vec{m}$ at the hand of numerical examples below.
If the bath initially contains only a finite number of particles, $\tr\bigl[(\id-\id_{\vec{m}})\hat{\varrho}_0\bigr]$ vanishes for appropriate $\vec m$. Such states include the vacuum state which is also the zero temperature thermal state for the particle mapping. For higher temperature thermal states of the bath, $\tr\bigl[(\id-\id_{\vec{m}})\hat{\varrho}_0\bigr]$ vanishes exponentially for large $\{m_i\}$. The total error induced on the expectation value of $\hat O$ due to (i) truncating the chain to finite length and (ii) the truncation of the local dimensions  is bounded by the sum of the two individual error bounds: $\Delta(t,L)+ \Delta_{\vec m}(t)$. This rigorously  bounds the error of approximating an infinite-dimensional bath of bosons by a chain of length $L$ made up of finite-dimensional subsystems with nearest neighbour interactions. If in addition we assume the system with Hamiltonian $\hat H_S$ to be a spin system, then the Hamiltonian is in the class which, as \cite{osbo06} shows, can be simulated with resources polynomial in $L$ and error $\epsilon$, and exponential in $|t|$.

\textit{Numerical example} -- As an example, we consider the spin-boson model with power-law spectral density,
\begin{equation}\label{SB spec}
J(\omega)=\pi\alpha\,\omega_{c}^{1-s}\omega^s \,\Theta(1-\omega/\omega_{c}),
\end{equation}
where $\Theta$ is the Heaviside step function.
This model has been extensively probed numerically, and there has been controversy over the accuracy of numerically derived critical exponents. One of the issues with the results was the inability to verify the local Fock space truncation errors \cite{Volt09,chin11}. The system Hamiltonian and interaction part are $\hat H_S=-\Delta\hat\sigma_x/2$ and $\hat A_S= \sigma_z/2$.
 The dissipation is known as Ohmic for $s=1$ and super ohmic for $s>1$. This can be written in the chain representation using Eq.~(\ref{particle rep}) (see the appendix for details). In Fig.~\ref{fig}, the bound for the particle mapping is plotted for the super-ohmic case and various $L$ and $\vec m$. Constants used for the simulation (see figure caption) are taken from the literature \cite{bull03}. The initial state of the bath corresponds to the zero temperature thermal state. We probe the same initial state for the case of ohmic dissipation and achieve qualitatively the same results (see appendix). Furthermore, we test the bound for a squeezed vacuum state of the bath, which is a highly populated state (see appendix). 

\textit{Conclusion} -- The detailed simulation of the interaction of a quantum system with a structured environments composed of harmonic oscillators has applications in a wide variety of scientific fields. The multitude of proposed algorithms to tackle this problem numerically lacked a method that delivers a simulation result with a rigorous error bound associated with it. In this work we derived error bounds that demonstrate that the recently developed TEDOPA can provide such a method. More specifically, obtaining Lieb--Robinson type expressions we provide complete error bounds on the simulation of observables of quantum systems coupled to a bosonic baths with infinitely many degrees of freedom such as the spin-boson model. This includes the errors incurred due to the truncation of the local Hilbert-spaces of the harmonic oscillators and due to the truncation of the length of the harmonic chain representing the environment. In this manner we provide a fully rigorous upper bound on the error for the numerical simulation of a spin-boson model and its generalisation to multiple baths and more general systems.

\textit{Acknowledgements} --
M.P.W. would like to thank Gerald Teschl for discussions regarding Jacobi operators and M.B.P. acknowledges discussions
with S.F. Huelga. This work was supported by the EPSRC CDT on Controlled Quantum Dynamics, the EU STREP projects PAPETS
and EQUAM, the EU Integrating project SIQS, and the ERC Synergy grant BioQ as well as the Alexander von Humboldt Foundation

\clearpage

\appendix
\widetext
\section{Spatial truncation of the bath}\label{spat bound}
We consider an arbitrary (not necessarily finite-dimensional) system described by a Hamiltonian $\hat{H}_S$ and a bosonic bath described by
\begin{equation}
\hat{H}_B=\frac{1}{2}\sum_{i,j=0}^\infty\bigl[\hat{x}_iX_{i,j}\hat{x}_j+\hat{p}_iP_{i,j}\hat{p}_j\bigr],
\end{equation}
where $\hat{x}_i$ and $\hat{p}_i$ are canonical position and momentum operators with the usual commutation relations $[\hat{x}_i,\hat{x}_j]=[\hat{p}_i,\hat{p}_j]=0$ and
$[\hat{x}_i,\hat{p}_j]=\mi\delta_{i,j}$. As we are allowing the bath to consist of infinitely-many modes (infinitely-many lattice sites),
we assume throughout that the domain of the Hamiltonian is well-defined. W.l.o.g., we let $X_{i,j}=X_{j,i}\in\rr$ and $P_{i,j}=P_{j,i}\in\rr$. We assume that they couple only nearest neighbours:
$X_{i,j}=P_{i,j}=0$ for $|i-j|>1$.

We suppose that system and bath are coupled according to
\begin{equation}
\hat{V}=\hat{h}\hat{x}_0,
\end{equation}
where  $\hat{h}$ acts on the system and we assume $\|\hat h\|<\infty$.
Thus, our total Hamiltonian reads
\begin{equation}
\begin{split}
\hat{H}=\hat{H}_S+\hat{V}+\hat{H}_B,
\end{split}
\end{equation}
where for compactness of notation, we have neglected tensor products with the identity. We are interested
in the error introduced for the time-evolution of bounded observables $\hat{O}$ (assuming $\|\hat{O}\|<\infty$)
acting on the system when, instead of simulating the full Hamiltonian $\hat{H}$, we take only finitely many
lattice sites of the bath into account. Namely those that are closest to the site $0$, truncating
the bath Hamiltonian according to
\begin{equation}
    \hat{H}_B^L=\frac{1}{2}\sum_{i,j=0}^{L-1}\bigl[\hat{x}_iX_{i,j}\hat{x}_j+\hat{p}_iP_{i,j}\hat{p}_j\bigr]
    =\frac{1}{2}\sum_{i,j}\bigl[\hat{x}_i(X_L)_{i,j}\hat{x}_j+\hat{p}_i(P_L)_{i,j}\hat{p}_j\bigr],
\end{equation}
where  $X_L$, $P_L$ are the principle submatrices of $X$, $P$ corresponding to the non-truncated modes. The
truncated chain hence consists of $L$ modes. Denoting the initial state of the whole system by $\hat{\varrho}_0$
and the total truncated Hamiltonian by $\hat{H}_L=\hat{H}_S+\hat{V}+\hat{H}^L_B$, we set out to bound the difference
\begin{equation}
    \Delta(t,L)=\bigl|\tr\bigl[\hat{O}\me^{-\mi\hat{H}t}\hat{\varrho}_0\me^{\mi\hat{H}t}\bigr] - \tr\bigl[\hat{O}\me^{-\mi\hat{H}_Lt}\hat{\varrho}_0\me^{\mi\hat{H}_Lt}\bigr]\bigr|.
\end{equation}
We will prove the following theorem.

\begin{theorem}[Spatial truncation of the bath]\label{bound 2} Let $\hat{H}$, $\hat{H}_L$ as above,
 $c$, $c^\prime$ such that $\|P_LX_L\|^{1/2}\le c$ and  $\max\{\|X\|,\|P\|\}\le c^\prime$. Then

\begin{equation}\label{alt bound}
\begin{split}
\Delta^2(t,L)&\le
4\|\hat{O}\|^2\frac{\|\hat{h}\|}{c}
\Bigl(\frac{\|P_L\||X_{L-1,L}|}{c^2}
+\frac{|P_{L-1,L}|}{c}
\Bigr)\frac{(ct)^{L+1}}{(L+1)!}(\me^{ct}+1)\Bigl(\|\gamma_0\|^{1/2}+\|\hat{h}\|
\frac{\me^{c^\prime t}-1}{c^\prime}\Bigr)\me^{c^\prime t}.
\end{split}
\end{equation}
If  $X,P>0$ or $X=P$, we may take $c^\prime\rightarrow 0$ such that we recover the theorem in the main text.
If $P\propto\id$, we may replace $\frac{(ct)^{1+L}}{(L+1)!}$ by $\frac{(ct)^{2L+1}}{(2L+1)!}$.
Here,
\begin{equation}
\gamma_0=\left(\begin{array}{cc}
\gamma_{xx} & \gamma_{xp}\\
\gamma_{xp}^\dagger & \gamma_{pp}
\end{array}\right),\;\;\; [\gamma_{xx}]_{i,j}=\tr[\hat{x}_i\hat{x}_j\hat{\varrho}_0],
\;\;\; [\gamma_{pp}]_{i,j}=\tr[\hat{p}_i\hat{p}_j\hat{\varrho}_0],\;\;\; [\gamma_{xp}]_{i,j}=\tr[\hat{x}_i\hat{p}_j\hat{\varrho}_0],
\end{equation}
collects the two-point bath correlations in the initial state of the whole system. Note that $\|X_L\| \leq \|X\|$ and $\|P_L\| \leq \|P\|$.
\end{theorem}

One can allow for the two-point correlations collected in $\gamma_0$ to diverge and still get a bound on $\Delta(t,L)$, see Section~\ref{infinite_correlations}.
Often, one encounters systems interacting with multiple baths. We generalize to this setting in Section~\ref{sec:mult_baths}.

\subsection{Proof}
Denote
\begin{equation}
\hat{U}(t)=\me^{\mi t(\hat{H}-\hat{V})}\me^{-\mi t\hat{H}},\;\;\;
\hat{U}_L(t)=\me^{\mi t(\hat{H}_L-\hat{V})}\me^{-\mi t\hat{H}_L}.
\end{equation}
Then for system operators $\hat O$
\begin{equation}
\label{same_start}
\begin{split}
\tr[\hat{O}\me^{-\mi t\hat{H}}\hat{\varrho}_0\me^{\mi t\hat{H}}]=\tr[\hat{O}\me^{-\mi t(\hat{H}-\hat{V})}\hat{U}(t)\hat{\varrho}_0\hat{U}^\dagger(t)\me^{\mi t(\hat{H}-\hat{V})}]=\tr[\me^{\mi t\hat{H}_S}\hat{O}\me^{-\mi t\hat{H}_S}\hat{U}(t)\hat{\varrho}_0\hat{U}^\dagger(t)]
\end{split}
\end{equation}
and similarly
\begin{equation}
\begin{split}
\tr[\hat{O}\me^{-\mi t\hat{H}_L}\hat{\varrho}_0\me^{\mi t\hat{H}_L}]=\tr[\me^{\mi t\hat{H}_S}\hat{O}\me^{-\mi t\hat{H}_S}\hat{U}_L(t)\hat{\varrho}_0\hat{U}_L^\dagger(t)].
\end{split}
\end{equation}
Hence,
\begin{equation}
\begin{split}
\tr[\hat{O}\me^{-\mi t\hat{H}}\hat{\varrho}_0\me^{\mi t\hat{H}}]-\tr[\hat{O}\me^{-\mi t\hat{H}_L}\hat{\varrho}_0\me^{\mi t\hat{H}_L}]
&=\tr\bigl[\me^{\mi t\hat{H}_S}\hat{O}\me^{-\mi t\hat{H}_S}\bigl(\hat{U}(t)\hat{\varrho}_0[\hat{U}^\dagger(t)-\hat{U}_L^\dagger(t)]+[\hat{U}(t)-\hat{U}_L(t)]\hat{\varrho}_0\hat{U}_L^\dagger(t)\bigr)\bigr].
\end{split}
\end{equation}
Using the Cauchy-Schwarz inequality $|\text{tr}[\hat{A}\hat{B}\hat{\varrho}_0]|^2\le \text{tr}[\hat{A}\hat{A}^\dagger\hat{\varrho}_0]\text{tr}[\hat{B}^\dagger\hat{B}\hat{\varrho}_0]\le \|\hat{A}\|^2\text{tr}[\hat{B}^\dagger\hat{B}\hat{\varrho}_0]$ and the triangle inequality, we find
\begin{equation}
\label{same_end}
\begin{split}
\Delta(t,L)=
\bigl|\tr[\hat{O}\me^{-\mi t\hat{H}}\hat{\varrho}_0\me^{\mi t\hat{H}}]-\tr[\hat{O}\me^{-\mi t\hat{H}_L}\hat{\varrho}_0\me^{\mi t\hat{H}_L}]\bigr|
&\le 2\|\hat{O}\|
\sqrt{\tr\bigl[[\hat{U}^\dagger(t)-\hat{U}_L^\dagger(t)][\hat{U}(t)-\hat{U}_L(t)]\hat{\varrho}_0\bigr]},
\end{split}
\end{equation}
where
\begin{equation}
\begin{split}
\tr\bigl[[\hat{U}^\dagger(t)-\hat{U}_L^\dagger(t)][\hat{U}(t)-\hat{U}_L(t)]\hat{\varrho}_0\bigr]&=-2\,\Re\int_0^t\md x\,\frac{\md}{\md x}\tr\bigl[\hat{U}^\dagger(x)\hat{U}_L(x)\hat{\varrho}_0\bigr]
\end{split}
\end{equation}
and
\begin{equation}
\begin{split}
-\mi
\frac{\md}{\md x}\hat{U}^\dagger(x)\hat{U}_L(x)&=
\hat{U}^\dagger(x)\me^{\mi x\hat{H}_S}\me^{\mi x\hat{H}_B}\bigl(
\hat{V}
-
\me^{-\mi x\hat{H}_B}\me^{\mi x\hat{H}_B^L}\hat{V}\me^{-\mi x\hat{H}^L_B}\me^{\mi x\hat{H}_B}\bigr)\me^{-\mi x\hat{H}_B}\me^{\mi x\hat{H}^L_B}\me^{-\mi x\hat{H}_L},
\end{split}
\end{equation}
where
\begin{equation}
\label{eq:up_to_here}
\begin{split}
\me^{-\mi x\hat{H}_B}\me^{\mi x\hat{H}_B^L}\hat{V}\me^{-\mi x\hat{H}^L_B}\me^{\mi x\hat{H}_B}-\hat{V}
&=-\mi\hat{h}\int_0^x\md y\,\me^{-\mi y\hat{H}_B}\left[(\hat{H}_B-\hat{H}_B^L),\me^{\mi y\hat{H}_B^L}\hat{x}_0\me^{-\mi y\hat{H}^L_B}\right]\me^{\mi y\hat{H}_B}.
\end{split}
\end{equation}
Let us summarize the bound so far:
\begin{equation}\label{so far}
\begin{split}
\frac{\Delta^2(t,L)}{8\|\hat{O}\|^2}&\le
\int_0^t\!\!\!\md x\!\int_0^x\!\!\!\!\md y\,
\bigl|
\tr\bigl[
\hat{U}^\dagger(x)\me^{\mi x\hat{H}_S}\me^{\mi x\hat{H}_B}
\me^{-\mi y\hat{H}_B}\hat{h}
\bigl[(\hat{H}_B-\hat{H}_B^L),\me^{\mi y\hat{H}_B^L}\hat{x}_0\me^{-\mi y\hat{H}^L_B}\bigr]\me^{\mi y\hat{H}_B}\me^{-\mi x\hat{H}_B}\me^{\mi x\hat{H}^L_B}\me^{-\mi x\hat{H}_L}
\hat{\varrho}_0\bigr]\bigr|.
\end{split}
\end{equation}
We now proceed to bound the commutator and come back to Eq.~(\ref{so far}) after Eq.~(\ref{final_commutator}).
We have
\begin{equation}
\begin{split}
\hat{H}_B-\hat{H}_B^L&=\frac{1}{2}\sum_{i=L}^\infty\sum_{j=0}^{L-1}\bigl[\hat{x}_iX_{i,j}\hat{x}_j+\hat{p}_iP_{i,j}\hat{p}_j\bigr]
+\frac{1}{2}\sum_{i=0}^\infty\sum_{j=L}^{\infty}\bigl[\hat{x}_iX_{i,j}\hat{x}_j+\hat{p}_iP_{i,j}\hat{p}_j\bigr]\\
\end{split}
\end{equation}
such that, as only nearest neighbours are coupled and $X$ and $P$ are symmetric,
\begin{equation}
\label{final_commutator}
\begin{split}
\left[(\hat{H}_B-\hat{H}_B^L),\me^{\mi y\hat{H}_B^L}\hat{x}_0\me^{-\mi y\hat{H}^L_B}\right]&=
\left[\hat{x}_{L-1},\me^{\mi y\hat{H}_B^L}\hat{x}_0\me^{-\mi y\hat{H}^L_B}\right]X_{L-1,L}\hat{x}_{L}
+\left[\hat{p}_{L-1},\me^{\mi y\hat{H}_B^L}\hat{x}_0\me^{-\mi y\hat{H}^L_B}\right]P_{L-1,L}\hat{p}_L\\
&=:C_{0,L-1}^{xx}(y)X_{L-1,L}\hat{x}_{L}
+C_{0,L-1}^{xp}(y)P_{L-1,L}\hat{p}_L.
\end{split}
\end{equation}
Let us now come back to Eq.~(\ref{so far}). Inserting the above expression, we see that we need to bound terms of the form
\begin{equation}
\begin{split}
F_r(x,y)=
\bigl|
\tr\bigl[
\hat{U}^\dagger(x)\me^{\mi x\hat{H}_S}\me^{\mi x\hat{H}_B}
\me^{-\mi y\hat{H}_B}\hat{h}
\hat{r}_L\me^{\mi y\hat{H}_B}\me^{-\mi x\hat{H}_B}\me^{\mi x\hat{H}^L_B}\me^{-\mi x\hat{H}_L}
\hat{\varrho}_0\bigr]\bigr|.
\end{split}
\end{equation}
with $r=x,p$. Writing
$\hat{r}_L(t)=\me^{\mi t\hat{H}_B}\hat{r}_L\me^{-\mi t\hat{H}_B}$, $\hat{\varrho}(x)=\me^{\mi x(\hat{H}_L-\hat{V})}\me^{-\mi x\hat{H}_L}
\hat{\varrho}_0\me^{\mi x\hat{H}_L}\me^{-\mi x(\hat{H}_L-\hat{V})}$, inserting the definition of $\hat{U}(x)$, and using $[\hat{H}_S,\hat{H}_B]=[\hat{H}_S,\hat{H}^L_B]=[\hat{H}_S,\hat{r}_L(t)]=0$, this reads
\begin{equation}
\label{single_F}
\begin{split}
F_r(x,y)&=
\bigl|
\tr\bigl[\me^{\mi x(\hat{H}_L-\hat{V})}\me^{-\mi x\hat{H}_L}
\me^{\mi x\hat{H}}
\hat{h}\me^{-\mi x\hat{H}_S}\me^{-\mi x\hat{H}_B}\hat{r}_L(x-y)\hat{\varrho}(x)\bigr]\bigr|
\le \|\hat{h}\|\sqrt{
\tr\bigl[\hat{r}^2_L(x-y)\hat{\varrho}(x)\bigr]},
\end{split}
\end{equation}
where we used $|\text{tr}[\hat{A}\hat{B}\hat{\varrho}(x)]|^2\le \|\hat{A}\|^2\text{tr}[\hat{B}^\dagger\hat{B}\hat{\varrho}(x)]$ to obtain the second line.
Inserting Eqs.~(\ref{final_commutator},\ref{single_F}) into  Eq.~(\ref{so far}), we hence have
\begin{equation}\label{F_before_LR}
\begin{split}
\frac{\Delta^2(t,L)}{8\|\hat{O}\|^2\|\hat{h}\|}&\le
|X_{L-1,L}|\int_0^t\!\!\!\md x\!\int_{0}^x\!\!\!\!\md y\,C^{xx}_{0,L-1}(x-y)\gamma_x(x,y)
+|P_{L-1,L}|\int_0^t\!\!\!\md x\!\int_{0}^x\!\!\!\!\md y\,C^{xp}_{0,L-1}(x-y)\gamma_p(x,y),
\end{split}
\end{equation}
where we denoted
\begin{equation}
\begin{split}
\gamma_r(x,y)&=\sqrt{\tr\bigl[\hat{r}^2_L(y)\hat{\varrho}(x)\bigr]}.
\end{split}
\end{equation}
To keep track of the case $P\propto \id$, we let $P_{i,j}=0$ for $|i-j|>R$ with $R=0,1$. By Eq.~(56) in Ref.~\cite{Marcus_bosonic_LR} and as $\|X_LP_L\|=\|P_LX_L\|$ \cite{horn90}, we have
\begin{equation}\label{final_LR_bounds}
\begin{split}
|C^{xx}_{0,L-1}(y)|&\le\sum_{\substack{n=0 \\ L\le (n+1)(1+R)}}^\infty\frac{|y|^{2n+1}}{(2n+1)!}\|P_LX_L\|^n\|P_L\|\\
|C_{0,L-1}^{xp}(y)|&\le \sum_{\substack{n=0 \\ L\le n(1+R)+1}}^\infty\frac{|y|^{2n}}{(2n)!}\|P_LX_L\|^n.
\end{split}
\end{equation}
Bounding the second moments $\gamma_r(x,y)$ in the following section, we return to Eq.~(\ref{F_before_LR}) in Section~\ref{appendix_final_steps} to complete the proof.

\subsubsection{Second moments}
\label{Gamma_bounds}
Recalling that $\hat{\varrho}(x)=\me^{\mi x(\hat{H}_L-\hat{V})}\me^{-\mi x\hat{H}_L}
\hat{\varrho}_0\me^{\mi x\hat{H}_L}\me^{-\mi x(\hat{H}_L-\hat{V})}$, we find ($r=x,y$)
\begin{equation}
\begin{split}
-\mi\frac{\partial}{\partial x}\tr\bigl[\hat{r}^2_L(y)\hat{\varrho}(x)\bigr]
&=\tr\bigl[\hat{r}^2_L(y)[\hat{\varrho}(x),\me^{\mi x(\hat{H}_L-\hat{V})}\hat{V}\me^{-\mi x(\hat{H}_L-\hat{V})}]\bigr]\\
&=\tr\bigl[\hat{\varrho}(x)\me^{\mi x\hat{H}_S}\hat{h}\me^{-\mi x\hat{H}_S}[\me^{\mi x\hat{H}_B^L}\hat{x}_0\me^{-\mi x\hat{H}_B^L},\hat{r}^2_L(y)]\bigr]\\
&=2\tr\bigl[\hat{\varrho}(x)\me^{\mi x\hat{H}_S}\hat{h}\me^{-\mi x\hat{H}_S}[\me^{\mi x\hat{H}_B^L}\hat{x}_0\me^{-\mi x\hat{H}_B^L},\hat{r}_L(y)]\hat{r}_L(y)\bigr].
\end{split}
\end{equation}
Now,
\begin{equation}
\label{the_symplectic_trafo}
\begin{split}
\hat{r}_k(y)=\me^{\mi y\hat{H}_B}\hat{r}_k\me^{-\mi y\hat{H}_B}=\sum_{l}c^{rx}_{k,l}(y)\hat{x}_l+\sum_lc^{rp}_{k,l}(y)\hat{p}_l,
\end{split}
\end{equation}
where $r=x,y$ and
\begin{equation}
\left(\begin{array}{cc}
c_{xx}(y) & c_{xp}(y)\\
c_{px}(y) & c_{pp}(y)
\end{array}
\right)=\me^{-\sigma H_By},\;\;\; H_B=X\oplus P, \;\;\;\sigma=\left(\begin{array}{cc}
0&-\id\\
\id &0
\end{array}\right).
\end{equation}
Hence,
\begin{equation}
\begin{split}
-\mi\frac{\partial}{\partial x}\tr\bigl[\hat{r}^2_L(y)\hat{\varrho}(x)\bigr]
&=
2\tr\bigl[\hat{\varrho}(x)\me^{\mi x\hat{H}_S}\hat{h}\me^{-\mi x\hat{H}_S}\hat{r}_L(y)\bigr]\sum_{l=0}^{L-1}\left([\me^{\mi x\hat{H}_B^L}\hat{x}_0\me^{-\mi x\hat{H}_B^L},c^{rx}_{L,l}(y)\hat{x}_l+c^{rp}_{L,l}(y)\hat{p}_l]\right)\\
&=
2\mi\tr\bigl[\hat{\varrho}(x)\me^{\mi x\hat{H}_S}\hat{h}\me^{-\mi x\hat{H}_S}\hat{r}_L(y)\bigr]\sum_{l=0}^{L-1}\left( c^{rp}_{L,l}(y)d^{xx}_{0,l}(x)-c^{rx}_{L,l}(y)d^{xp}_{0,l}(x)\right),
\end{split}
\end{equation}
where
\begin{equation}
\left(\begin{array}{cc}
d_{xx}(x) & d_{xp}(x)\\
d_{px}(x) & d_{pp}(x)
\end{array}
\right)=\me^{-\sigma_L H^L_Bx},\;\;\; H^L_B=X_L\oplus P_L,\;\;\;\sigma_L=\left(\begin{array}{cc}
0&-\id_L\\
\id_L &0
\end{array}\right).
\end{equation}
We find
\begin{equation}
\begin{split}
\me^{-\sigma H_By}\sigma\left[\left(\begin{array}{cc}
\id_L & 0 \\
0 & 0\\
0&  \id_L\\
0 & 0 \\
\end{array}
\right)\me^{-\sigma_L H^L_Bx}\left(\begin{array}{cccc}
\id_L & 0 & 0 &0\\
0 & 0 & \id_L & 0\\
\end{array}
\right)\right]^t
&=\left(\begin{array}{cc}
c_{xp}(y)\left(\begin{array}{cc}
d^t_{xx}(x)& 0\\
0 & 0  \\
\end{array}\right) -c_{xx}(y)\left(\begin{array}{cc}
d^t_{xp}(x) & 0\\
0 & 0  \\
\end{array}\right)& *\\
c_{pp}(y)\left(\begin{array}{cc}
d_{xx}^t(x)& 0\\
0 & 0  \\
\end{array}\right) -c_{px}(y)\left(\begin{array}{cc}
d^t_{xp}(x)  &0\\
0 & 0  \\
\end{array}\right)
& *
\end{array}
\right),
\end{split}
\end{equation}
the operator norm (and therefore the absolute value of all entries) of which is upper bounded by $ \|\me^{H_B\sigma y}\|\|\me^{H^L_B\sigma_L x}\|$.
Therefore, employing $|\text{tr}[\hat{A}\hat{B}\hat{\varrho}(x)]|^2\le \|\hat{A}\|^2\text{tr}[\hat{B}^\dagger\hat{B}\hat{\varrho}(x)]$,
\begin{equation}
\begin{split}
\Bigl|\frac{\partial}{\partial x}\tr\bigl[\hat{r}^2_L(y)\hat{\varrho}(x)\Bigr]\bigr|
&\le
2 \|\me^{H_B\sigma y}\|\|\me^{H^L_B\sigma_L x}\|\tr\bigl[\hat{\varrho}(x)\me^{\mi x\hat{H}_S}\hat{h}\me^{-\mi x\hat{H}_S}\hat{r}_L(y)\bigr]
\le 2 \|\me^{H_B\sigma y}\|\|\me^{H^L_B\sigma_L x}\|\|\hat{h}\|\sqrt{\tr\bigl[\hat{\varrho}(x)\hat{r}^2_L(y)\bigr] },
\end{split}
\end{equation}
which implies\footnote{
Suppose
\begin{equation}
\bigl|\frac{\md}{\md x}\alpha(x)\bigr|\le f(x)\sqrt{\alpha(x)}
\end{equation}
and let $\alpha(0)=\alpha_0\ge 0$. Then
\begin{equation}
\begin{split}
\alpha(x)&=\alpha_0+\int_0^x\!\!\!\md y\,\frac{\md}{\md y}\alpha(y)
\le\alpha_0+\int_0^x\!\!\!\md y\,f(y)\sqrt{\alpha(y)}=:\beta(x).
\end{split}
\end{equation}
Let $\epsilon>0$ and define $\gamma(x)=\beta(x)+\epsilon$. Then $\gamma(x)\ge\gamma(0)=a_0+\epsilon>0$, i.e., $\sqrt{\gamma}$ is differentiable and
\begin{equation}
\begin{split}
\sqrt{\beta(x)+\epsilon}&=
\sqrt{\gamma(x)}=\sqrt{\alpha_0+\epsilon}+\int_0^x\!\!\!\md y\,\frac{\md}{\md y}\sqrt{\gamma(y)},
\end{split}
\end{equation}
where
\begin{equation}
\begin{split}
\int_0^x\!\!\!\md y\,\frac{\md}{\md y}\sqrt{\gamma(y)}=\frac{1}{2}\int_0^x\!\!\!\md y\,\frac{f(y)\sqrt{\alpha(y)}}{\sqrt{\gamma(y)}}\le\frac{1}{2}\int_0^x\!\!\!\md y\,f(y).
\end{split}
\end{equation}
As $\epsilon>0$ was arbitrary, we hence have
\begin{equation}
\sqrt{\alpha(x)}\le \sqrt{\alpha}_0+\frac{1}{2}\int_0^x\md y\,f(y).
\end{equation}
}
\begin{equation}
\begin{split}
\gamma_r(x,y)\le \sqrt{\tr\bigl[\hat{r}^2_L(y)\hat{\varrho}_0\bigr]}+\|\hat{h}\|\|\me^{H_B\sigma y}\|\int_0^x\md z\,\|\me^{H^L_B\sigma_L z}\|.
\end{split}
\end{equation}
From Eq.~(\ref{the_symplectic_trafo}), we have, denoting $[\gamma_{xx}(y)]_{i,j}=\tr[\hat{x}_i(y)\hat{x}_j(y)\hat{\varrho}_0]$, $[\gamma_{xp}(y)]_{i,j}=\tr[\hat{x}_i(y)\hat{p}_j(y)\hat{\varrho}_0]$, $[\gamma_{px}(y)]_{i,j}=\tr[\hat{p}_i(y)\hat{x}_j(y)\hat{\varrho}_0]$, $[\gamma_{pp}(y)]_{i,j}=\tr[\hat{p}_i(y)\hat{p}_j(y)\hat{\varrho}_0]$,
\begin{equation}
\label{corr_of_y}
\begin{split}
\left(\begin{array}{cc}
\gamma_{xx}(y) & \gamma_{xp}(y)\\
\gamma_{px}(y) & \gamma_{pp}(y)
\end{array}
\right)&=\me^{-\sigma H_By}
\gamma_0(\me^{-\sigma H_By})^t,
\end{split}
\end{equation}
i.e., all entries are bounded from above by $\|\gamma_0\| \|\me^{H_B\sigma y}\|^2$, in particular $[\gamma_{rr}(y)]_{L,L}$ such that
\begin{equation}
\label{correlation_bound}
\begin{split}
\gamma_r(x,y)\le \|\gamma_0\|^{1/2} \|\me^{H_B\sigma y}\|+\|\hat{h}\|\|\me^{H_B\sigma y}\|\int_0^x\md z\,\|\me^{H^L_B\sigma_L z}\|.
\end{split}
\end{equation}
If $H_B>0$, we may use the Williamson normal form to write $ H_B\sigma= S^t(D\oplus D)S\sigma y=S^t(D\oplus D)\sigma (S^t)^{-1} y$, where
$ (D\oplus D)\sigma$ is real skew-symmetric, i.e., its eigenvalues are purely imaginary. Hence,
$\|\me^{ H_B \sigma y}\|=\|\me^{ (D\oplus D)\sigma y}\|=1$. We also have $\|\me^{ H_B \sigma y}\|=1$ if $X=P$ as then
$\sigma H_B$ is real skew-symmetric. Hence, $c^\prime$ an upper bound to $\max\{\|X\|,\|P\|\}$, we have
\begin{equation}
\label{final_Gamma_bound}
\begin{split}
\gamma_r(x,y)&\le
\begin{cases}
\|\gamma_0\|^{1/2}+\|\hat{h}\|x & \text{ if } X,P>0 \text{ or } X=P,\\
 \|\gamma_0\|^{1/2} \me^{c^\prime |y|}+\|\hat{h}\|\me^{c^\prime |y|}\frac{\me^{c^\prime x}-1}{c^\prime} & \text{ otherwise,}
\end{cases}\\
&=:\gamma(x,y).
\end{split}
\end{equation}

\subsubsection{Final steps}
\label{appendix_final_steps}
Inserting the bounds in Eq.~(\ref{final_LR_bounds}) and Eq.~(\ref{final_Gamma_bound})
into Eq.~(\ref{F_before_LR}) and letting $c$ such that $\sqrt{\|P_LX_L\|}\le c$, we have
\begin{equation}
\label{blahblahblah}
\begin{split}
\frac{\Delta^2(t,L)}{8\|\hat{O}\|^2\|\hat{h}\|}&\le
\int_0^t\!\!\!\md x\!\int_{0}^x\!\!\!\!\md y\,\gamma(x,x-y)\Bigl(\frac{\|P_L\||X_{L-1,L}|}{c}\sum_{n=\lceil\frac{L}{1+R}\rceil }^\infty\frac{(cy)^{2n-1}}{(2n-1)!}
+|P_{L-1,L}|\sum_{n=\lceil\frac{L-1}{2}\rceil}^\infty\frac{(cy)^{2n}}{(2n)!}\Bigr),
\end{split}
\end{equation}
where
\begin{equation}
\begin{split}
\int_0^t\!\!\!\md x\,\int_{0}^x\!\!\!\!\md y\,\gamma(x,x-y)y^n&=\|\gamma_0\|^{1/2} \int_0^t\!\!\!\md x\,\int_{0}^x\!\!\!\!\md y\,\me^{c^\prime (x-y)}y^n+\|\hat{h}\|\int_0^t\!\!\!\md x\,\int_{0}^x\!\!\!\!\md y\,\me^{c^\prime (x-y)}\frac{\me^{c^\prime x}-1}{c^\prime}y^n\\
&\le \frac{t^{n+2}}{(n+1)(n+2)}\Bigl(\|\gamma_0\|^{1/2}+\|\hat{h}\|
\frac{\me^{c^\prime t}-1}{c^\prime}\Bigr)\me^{c^\prime t}
\end{split}
\end{equation}
and we may take $c^\prime\rightarrow 0$ if $X,P>0$ or $X=P$.
For $L$ even (odd) we have $\lceil\frac{L}{2}\rceil=L/2$ ($\lceil\frac{L}{2}\rceil=\frac{L+1}{2}$)  and $\lceil\frac{L-1}{2}\rceil=L/2$ ($\lceil\frac{L-1}{2}\rceil=\frac{L-1}{2}$). Hence, for $R=1$ the bound in the theorem follows.
Finally, for $P\propto\id$ the second term in Eq.~\eqref{blahblahblah} vanishes and we have $R=0$.

\subsubsection{Correlation matrix}
\label{infinite_correlations}
The upper bound on the correlations $\gamma_{ab}(y)$, $a,b\in\{x,p\}$, in Eq.~(\ref{correlation_bound}) may be altered to allow for divergences at infinity by recalling that
(see Eqs.~(\ref{the_symplectic_trafo},\ref{corr_of_y}))
\begin{equation}
\begin{split}
\left(\begin{array}{cc}
\gamma_{xx}(y) & \gamma_{xp}(y)\\
\gamma_{px}(y) & \gamma_{pp}(y)
\end{array}
\right)&=\me^{-\sigma H_By}
\gamma_0(\me^{-\sigma H_By})^t=\left(\begin{array}{cc}
c_{xx}(y) & c_{xp}(y)\\
c_{px}(y) & c_{pp}(y)
\end{array}
\right)\gamma_0\left(\begin{array}{cc}
c^t_{xx}(y) & c^t_{px}(y)\\
c^t_{xp}(y) & c^t_{pp}(y)
\end{array}
\right),
\end{split}
\end{equation}
i.e. (see Eq.~(\ref{corr_of_y})),
\begin{equation}
\begin{split}
\gamma_{xx}(y)&=c_{xx}(y)[\gamma_{xx}(0)c^t_{xx}(y) + \gamma_{xp}(0)c^t_{xp}(y)] + c_{xp}(y)[\gamma_{px}(0)c^t_{xx}(y) + \gamma_{pp}(0)c^t_{xp}(y) ], \\
\gamma_{pp}(y)&=c_{px}(y)[\gamma_{xx}(0)c^t_{px}(y) + \gamma_{xp}(0)c^t_{pp}(y)] + c_{pp}(y)[\gamma_{px}(0)c^t_{px}(y) + \gamma_{pp}(0)c^t_{pp}(y)],
\end{split}
\end{equation}
and again using the bounds obtained in \cite{Marcus_bosonic_LR} which, however, increases the value of $c$ in the bound.

\subsubsection{Multiple baths}
\label{sec:mult_baths}
For some applications in quantum biology and condensed matter physics, one has a quantum system coupled to $N$ baths which, using the Particle or Phonon mapping, can be written in the form
\be\label{the m bathtub}
\hat H^{mul.}=\hat H_S+\sum_{m=1}^N \hat H^{(m)}+\sum_{m=1}^N\hat h^{(m)} \hat x_0^{(m)}
\ee
where
\be
\hat H^{(m)}=\frac{1}{2}\sum_{i,j=0}^\infty \left[ \hat x_i^{(m)} X_{i,j}^{(m)} \hat x_j^{(m)}+\hat p_i^{(m)} P_{i,j}^{(m)}\hat p_j^{(m)} \right],
\ee
and $X_{i,j}^{(m)}=X_{j,i}^{(m)}\in\rr$, $P_{i,j}^{(m)}=P_{j,i}^{(m)}\in\rr$. As in the rest of this work so far we assume that they couple only nearest neighbours, i.e. $X_{i,j}^{(m)}=P_{i,j}^{(m)}=0$ for $|i-j|>1$.
We can truncate the $N$ chains such that the $m^\textup{th}$ chain contains $L_m$ modes:
\begin{equation}\label{the m bathtub trunc}
\hat H_{\vec L}^{mul.} =\hat H_S+\sum_{m=1}^N \hat H^{(m)}_{L_m}+\sum_{m=1}^N\hat h^{(m)} \hat x_0^{(m)},
\end{equation}
where
\be
\hat H^{(m)}_{L_m}=\frac{1}{2}\sum_{i,j=0}^{L_m-1} \left[ \hat x_i^{(m)} X_{i,j}^{(m)} \hat x_j^{(m)}+\hat p_i^{(m)} P_{i,j}^{(m)}\hat p_j^{(m)} \right]=\frac{1}{2}\sum_{i,j} \left[ \hat x_i^{(m)} \left(X_{L_m}\right)_{i,j} \hat x_j^{(m)}+\hat p_i^{(m)}\left(P_{L_m}\right)_{i,j}\hat p_j^{(m)} \right],
\ee
where $X_{L_m}$ and $P_{L_m}$ are principle submatrices of $X^{(m)}$ and $P^{(m)}$ corresponding to the non-truncated modes. These definitions are in analogy with those at the beginning of section \ref{spat bound} but generalised to the case of $N$ non identical copies of the bath.
\begin{corollary}[Multiple chains]
Let $\hat H^{mul.}$, $\hat H^{mul.}_{\vec L}$ as above, $c_m,c_m^{\prime}$ such that $\|P_{L_m} X_{L_m}\|^{1/2}\leq c_m$ and $\textup{max}\left\{ \|X^{(m)}\|,\|P^{(m)}\|\right\}$ $\leq c_m'$. Then the error in truncating $\hat H^{mul.}$ by $\hat H^{mul.}_{\vec L}$ is bounded by
\begin{equation}
\Delta(\vec L, t):= \bigl|\tr\bigl[\hat{O}\me^{-\mi\hat{H^{mul.}}t}\hat{\varrho}_0\me^{\mi\hat{H^{mul.}}t}\bigr]-\tr\bigl[\hat{O}\me^{-\mi\hat H^{mul.}_{\vec L}t}\hat{\varrho}_0\me^{\mi\hat H^{mul.}_{\vec L}t}\bigr]\bigr|  \leq \sum_{m=1}^N F(m,t,L_m)
\end{equation}
where we have defined $F\geq 0$ as
\be
    F^2(m,t,L):= 4\|\hat{O}\|^2\frac{\|\hat{h}^{(m)}\|}{c_m}
    \Bigl(\frac{\|P_{L}\||X^{(m)}_{L-1,L}|}{c_m^2}
    +\frac{|P_{L-1,L}^{(m)}|}{c_m}
    \Bigr)\frac{(c_m t)^{L+1}}{(L+1)!}(\me^{c_m t}+1)\Bigl(\|\gamma_0^{(m)}\|^{1/2}+\|\hat{h}^{(m)}\|
    \frac{\me^{c_m^\prime t}-1}{c_m^\prime}\Bigr)\me^{c_m^\prime t}.
\ee
If  $X^{(m)},P^{(m)}>0$ or $X^{(m)}=P^{(m)}$, we may take $c_m^\prime\rightarrow 0$.
If $P^{(m)}\propto\id$, we may replace $\frac{(c_mt)^{1+L_m}}{(L_m+1)!}$ by $\frac{(c_m t)^{2L_m+1}}{(2L_m+1)!}$.
Here,
\begin{equation}
\gamma_0^{(m)}=\left(\begin{array}{cc}
\gamma_{xx}^{(m)} & \gamma_{xp}^{(m)}\\
\gamma_{xp}^{(m)^\dagger} & \gamma_{pp}^{(m)}
\end{array}\right),\;\;\; [\gamma_{xx}^{(m)}]_{i,j}=\tr[\hat{x}^{(m)}_i\hat{x}^{(m)}_j\hat{\varrho}_0],
\;\;\; [\gamma^{(m)}_{pp}]_{i,j}=\tr[\hat{p}^{(m)}_i\hat{p}^{(m)}_j\hat{\varrho}_0],\;\;\; [\gamma^{(m)}_{xp}]_{i,j}=\tr[\hat{x}^{(m)}_i\hat{p}^{(m)}_j\hat{\varrho}_0],
\end{equation}
collects the two-point $m^\text{th}$ bath correlations in the initial  state of the whole system. Note that $\|X_{L_m}^{(m)}\| \leq \|X^{(m)}\|$ and $\|P_{L_m}^{(m)}\| \leq \|P^{(m)}\|$.
\end{corollary}
As with theorem \ref{bound 2}, one can allow for the two-point correlations collected in $\gamma_0^{(m)}$ to diverge and still get a bound on $F$, see Section~\ref{infinite_correlations}.
Often, one encounters systems interacting with multiple baths. We generalize to this setting in Section~\ref{sec:mult_baths}.
\proof{
Starting from
\begin{equation}
\tr\bigl[\hat{O}\me^{-\mi\hat{H^{mul.}}t}\hat{\varrho}_0\me^{\mi\hat{H^{mul.}}t}\bigr]-\tr\bigl[\hat{O}\me^{-\mi\hat H^{mul.}_{\vec L}t}\hat{\varrho}_0\me^{\mi\hat H^{mul.}_{\vec L}t}\bigr],
\end{equation}
 we add and subtract
\begin{equation}
\tr\bigl[\hat{O}\me^{-\mi\hat{H}_{tru.}t}\hat{\varrho}_0\me^{\mi\hat{H}_{tru.}t}\bigr]
\end{equation}
$N-1$ times where $\hat{H}_{tru.}$ corresponds to $H^{mult.}$ but with some of the $N$ baths truncated. Each time it is added and subtracted, different baths should be truncated. We then group the terms in pairs of 2 and redefine the system in each pair such that the system contains $N-1$ baths (some truncated, some not). This step relies crucially on the fact that the system Hamiltonian $\hat H_s$ in not necessarily bounded. We then take the absolute value and apply the triangle inequality to the pairs followed by applying theorem \ref{theorem 1} to each pair.}

Thus the error introduced by truncating $N$ chains, is bounded by the sum of the errors of truncating each chain individually. The explicit forms of the bound for the Particle and Phonon mapping can be found in section \ref{sec:Derivation of the particle and phonon mapping chain truncation bounds}.

\section{Derivation of the particle and phonon mapping chain truncation bounds}\label{sec:Derivation of the particle and phonon mapping chain truncation bounds}
From \cite{misc14} we find Particle and Phonon mappings of Eq. \eqref{the bathtub}  to be
\begin{align}\label{particle rep actual}
\begin{split}
\hat H=&\hat H_S+\sqrt{\beta_0(0)}\hat A_S(b_0(0)+b_0^\dagger(0))
+\sum_{n=0}^\infty \Big(\alpha_n(0)b_n^\dagger(0) b_n(0)
+\sqrt{\beta_{n+1}(0)}(b_{n+1}^\dagger(0) b_n(0) +h.c.)\Big)
\end{split}
\end{align}
and
\begin{align}\label{phonon rep actual}
\begin{split}
\hat H=& \hat H_S+\sqrt{\beta_0(1)}\hat A_S \hat x_0(1)+\sum_{n=0}^\infty\Big(  \frac{\alpha_n(1)}{2} \hat x_n^2(1)
+\frac{1}{2}\hat p_n^2(1) +\sqrt{\beta_{n+1}(1)}\hat x_n(1) \hat x_{n+1}(1)\Big),
\end{split}
\end{align}
respectively. $b_n^\dagger(0),$ ($b_n(0)$), are creation (annihilation) operators. Define position and momentum operators for the particle mapping $\hat x_n(0):=(b^\dagger_n(0)+b_n(0))/\sqrt{2},$ $\hat p_n(0):=\mi(b^\dagger_n(0)-b_n(0))/\sqrt{2}$ and $\underline {\hat x}_n(1):=\sqrt{\omega_{max}}\hat x_n(1),$ $\underline {\hat p}_n(1):=\hat p_n(1)/\sqrt{\omega_{max}}$ for the phonon mapping. 
Write Eqs. (\ref{particle rep actual}), (\ref{phonon rep actual}) in terms of these new operators and compare these Eqs. with  Eqs. (\ref{the bath}) and (\ref{tot ham}). From here, together with the definition of the Jacobi matrices $\mathcal{J}(d\lambda^q)$ (see Eq. (162) in \cite{misc14}), we find:\\
For the particle mapping
\be\label{PM X P Jacobi rel}
X=P=\mathcal{J}(d\lambda^0),\quad \hat h=\sqrt{2\beta_0(0)}\hat A_S,\quad d\lambda^0(x)=J(x)dx/\pi.
\ee
For the phonon mapping
\be
X=\frac{\mathcal{J}(d\lambda^1)}{\omega_{max}},\quad P=\id\;\omega_{max}, \quad\hat h=\sqrt{\frac{\beta_0(1)}{\omega_{max}}}\hat A_S,\quad d\lambda^1(x)=J(\sqrt{x})dx/\pi.\label{ph map dens}
\ee
From Eqs (15,156,160) in \cite{misc14},
\be
\beta_0(0)=\int_{\omega_{min}}^{\omega_{max}} dxJ(x)/\pi,\quad \beta_0(1)=\int_{\omega^2_{min}}^{\omega^2_{max}} dx J(\sqrt{x})/\pi.
\ee
Since the spectrum of a Jacobi matrix is equal to its minimally closed support interval \cite{jacobi}, we have for the particle and phonon mappings: $\|X\|=\|P\|=\sqrt{\|XP\|}=\omega_{max}$, and $X>0$ iff $\omega_{min}>0$. For the Particle mapping we can use Eq. (\ref{alt bound}) with $c= c'= \omega_{max}$ to achieve
\be\label{eq:particle mapping boudn result}
\Delta^2(t,L)\leq 8 \mu_0 \|\hat O\|^2 \frac{\|\hat A_S \|}{\omega_{max}}\frac{(\omega_{max} t)^{L+1}}{(L+1)!} \left( \me^{\omega_{max} t}+1 \right)\left( \|\gamma_0\|^{1/2}+\mu_0\|\hat A_S\| t  \right)
\ee
where $\mu_0$ given by Eq. \eqref{eq: mu 1 and mu 2}. For the massive Phonon mapping, we replace
$8$ with $4$, $\mu_0$ with $\mu_1$, and $(\omega_{max} t)^{L+1}/(L+1)!$ by $(\omega_{max} t)^{2L+1}/(2L+1)!$ in Eq. \eqref{eq:particle mapping boudn result}.
For the massless Phonon chain mapping, we use Eq. (\ref{alt bound}) again, to achieve
\be\label{eq:phonon mapping boudn result}
\Delta^2(t,L)\leq 4 \mu_1 \|\hat O\|^2 \frac{\|\hat A_S \|}{\omega_{max}}\frac{(\omega_{max} t)^{2L+1}}{(2L+1)!} \left( \me^{\omega_{max} t}+1 \right)\left( \|\gamma_0\|^{1/2}+\|\hat h\| \frac{\me^{\omega_{max}t}-1}{\omega_{max}}  \right)\me^{\omega_{max} t}.
\ee
We can write the $\gamma_0$ matrix for the Phonon mapping in terms of the original $\hat x_n$ and $\hat p_n$ coordinates of Eq. \ref{phonon rep}, to find
\begin{equation}
\gamma_0=\left(\begin{array}{cc}
\omega_{max}\gamma_{xx} & \gamma_{xp}\\
\gamma_{px} & \frac{1}{\omega_{max}}\gamma_{pp}
\end{array}\right),\;\;\; [\gamma_{ab}]_{n,l}=\tr[\hat{a}_n\hat{b}_l\hat{\varrho}_0].
\end{equation}

\section{Fock space truncation}
In this section we derive bounds on the error introduced by truncating the local Hilbert spaces of the bath.
To this end, we define the projector
\begin{equation}
\id_{\vec{m}}=\id_{m_0}\otimes\id_{m_1}\otimes\cdots,
\end{equation}
where $\id_{m_i}$ acts on the $i$'th site of the bath and truncates the local Hilbert space according to
\begin{equation}
\id_{m}=\sum_{n=0}^{m}|n\rangle\langle n|.
\end{equation}
For bounded observables acting on the system $\hat{O}$, $\|\hat{O}\|<\infty$, we consider
\begin{equation}
\Delta_{\vec{m}}(t)=
\bigl|\tr[\hat{O}\me^{-\mi t\hat{H}}\hat{\varrho}_0\me^{\mi t\hat{H}}]-\tr[\hat{O}\me^{-\mi t\hat{H}_{\vec{m}}}\hat{\varrho}_0\me^{\mi t\hat{H}_{\vec{m}}}]\bigr|,
\end{equation}
i.e., the error
introduced by evolving the system according to
\begin{equation}
\hat{H}_{\vec{m}}=\id_{\vec{m}}\hat{H}\id_{\vec{m}}=\hat{H}_S+\hat{H}_B^{\vec{m}}+\hat{V}_{\vec{m}}
\end{equation}
instead of $\hat{H}=\hat{H}_S+\hat{H}_B+\hat{V}$. Here, with the notation $\hat{x}_i^{\vec{m}}=\id_{\vec{m}}\hat{x}_i\id_{\vec{m}}$ and $\hat{p}_i^{\vec{m}}=\id_{\vec{m}}\hat{p}_i\id_{\vec{m}}$, the individual terms read
 $\hat{V}_{\vec{m}}=\hat{h}\otimes\hat{x}^{\vec{m}}_0$ and
 \begin{equation}
 \begin{split}
 \hat{H}_B^{\vec{m}}&=\id_{\vec{m}}\hat{H}_B\id_{\vec{m}}=
 \frac{1}{2}\sum_{i,j}\bigl[X_{i,j}\id_{\vec{m}}\hat{x}_i\hat{x}_j\id_{\vec{m}}+P_{i,j}\id_{\vec{m}}\hat{p}_i\hat{p}_j\id_{\vec{m}}\bigr],
 \end{split}
 \end{equation}
where we note that $\id_{\vec{m}}\hat{x}_i^2\id_{\vec{m}}\ne (\hat{x}^{\vec{m}}_i)^2$ and
$\id_{\vec{m}}\hat{p}_i^2\id_{\vec{m}}\ne (\hat{p}^{\vec{m}}_i)^2$, while for $i\ne j$ we do have
$\id_{\vec{m}}\hat{x}_i\hat{x}_j\id_{\vec{m}}=\hat{x}^{\vec{m}}_i\hat{x}^{\vec{m}}_j$ and
$\id_{\vec{m}}\hat{p}_i\hat{p}_j\id_{\vec{m}}=\hat{p}^{\vec{m}}_i\hat{p}^{\vec{m}}_j$.
Now denote
\begin{equation}
\begin{split}
\hat{U}(t)&=\me^{\mi t(\hat{H}-\hat{V})}\me^{-\mi t\hat{H}},\\
\hat{U}_{\vec{m}}(t)&=\me^{\mi t(\hat{H}_{\vec{m}}-\hat{V}_{\vec{m}})}\me^{-\mi t\hat{H}_{\vec{m}}}.
\end{split}
\end{equation}
Proceeding as in Eqs. (\ref{same_start}-\ref{same_end}), we find
\begin{equation}
\begin{split}
\Delta_{\vec{m}}(t)=
\bigl|\tr[\hat{O}\me^{-\mi t\hat{H}}\hat{\varrho}_0\me^{\mi t\hat{H}}]-\tr[\hat{O}\me^{-\mi t\hat{H}_{\vec{m}}}\hat{\varrho}_0\me^{\mi t\hat{H}_{\vec{m}}}]\bigr|
&\le 2\|\hat{O}\|
\sqrt{\tr\bigl[[\hat{U}^\dagger(t)-\hat{U}_{\vec{m}}^\dagger(t)][\hat{U}(t)-\hat{U}_{\vec{m}}(t)]\hat{\varrho}_0\bigr]},
\end{split}
\end{equation}
where now, as $\hat{U}_{\vec{m}}^\dagger(t)\hat{U}_{\vec{m}}(t)=\id_S\otimes\id_{\vec{m}}$,
\begin{equation}
\begin{split}
\tr\bigl[[\hat{U}^\dagger(t)-\hat{U}_{\vec{m}}^\dagger(t)][\hat{U}(t)-\hat{U}_{\vec{m}}(t)]\hat{\varrho}_0\bigr]&=
\tr\bigl[(\id-\id_{\vec{m}})\hat{\varrho}_0\bigr]-2\,\Re
\int_0^t\md x\,\frac{\md}{\md x}\tr\bigl[\hat{U}^\dagger(x)\hat{U}_{\vec{m}}(x)\hat{\varrho}_0\bigr]
\end{split}
\end{equation}
and
\begin{equation}
\begin{split}
-\mi
\frac{\md}{\md x}\hat{U}^\dagger(x)\hat{U}_{\vec{m}}(x)&=
\hat{U}^\dagger(x)\me^{\mi x\hat{H}_S}\hat{h}\bigl(\me^{\mi x\hat{H}_B}
\hat{x}_0\me^{-\mi x\hat{H}_B}
-
\me^{\mi x\hat{H}_B^{\vec{m}}}\hat{x}_0\me^{-\mi x\hat{H}_B^{\vec{m}}}\bigr)\me^{\mi x\hat{H}_B^{\vec{m}}}\me^{-\mi x\hat{H}_{\vec{m}}}\\
&=:\hat{U}^\dagger(x)\me^{\mi x\hat{H}_S}\hat{h}\hat{W}(x)\me^{\mi x\hat{H}_B^{\vec{m}}}\me^{-\mi x\hat{H}_{\vec{m}}}.
\end{split}
\end{equation}
The Cauchy--Schwarz inequality yields
\begin{equation}
\begin{split}
\bigl|\frac{\md}{\md x}\tr\bigl[\hat{U}^\dagger(x)\hat{U}_{\vec{m}}(x)\hat{\varrho}_0\bigr]\bigr|^2\le
\tr\bigl[\hat{h}^2
\me^{-\mi x\hat{H}_B^{\vec{m}}}\hat{W}^2(x)
\me^{\mi x\hat{H}_B^{\vec{m}}}\me^{-\mi x\hat{H}_{\vec{m}}}\hat{\varrho}_0\me^{\mi x\hat{H}_{\vec{m}}}\bigr]
=:\epsilon_{\vec{m}}(x)
\end{split}
\end{equation}
such that
\begin{equation}
\begin{split}
\Delta^2_{\vec{m}}(t)\le 4\|\hat{O}\|^2
\left(
\tr\bigl[(\id-\id_{\vec{m}})\hat{\varrho}_0\bigr]+2
\int_0^t\md x\,\sqrt{\epsilon_{\vec{m}}(x)}
\right).
\end{split}
\end{equation}
The error $\epsilon_{\vec{m}}(x)$ may be obtained numerically: We have
\begin{equation}
\begin{split}
\me^{-\mi x\hat{H}_B^{\vec{m}}}\hat{W}^2(x)
\me^{\mi x\hat{H}_B^{\vec{m}}}&=\me^{-\mi x\hat{H}_B^{\vec{m}}}\hat{W}(x)\id_{\vec{m}}\hat{W}(x)
\me^{\mi x\hat{H}_B^{\vec{m}}}+\me^{-\mi x\hat{H}_B^{\vec{m}}}\hat{W}(x)(\id-\id_{\vec{m}})\hat{W}(x)
\me^{\mi x\hat{H}_B^{\vec{m}}}\\
&=
\me^{-\mi x\hat{H}_B^{\vec{m}}}\bigl(
\hat{x}_0(x)
-
\me^{\mi x\hat{H}_B^{\vec{m}}}\hat{x}_0\me^{-\mi x\hat{H}_B^{\vec{m}}}\bigr)\id_{\vec{m}}\bigl(
\hat{x}_0(x)
-
\me^{\mi x\hat{H}_B^{\vec{m}}}\hat{x}_0\me^{-\mi x\hat{H}_B^{\vec{m}}}\bigr)
\me^{\mi x\hat{H}_B^{\vec{m}}}
\\
&\hspace{2cm}+\me^{-\mi x\hat{H}_B^{\vec{m}}}
\hat{x}_0(x)
(\id-\id_{\vec{m}})
\hat{x}_0(x)\me^{\mi x\hat{H}_B^{\vec{m}}}
\end{split}
\end{equation}
such that, recalling Eq.~\eqref{the_symplectic_trafo}, i.e., that $\hat{x}_0(t)=\me^{\mi t\hat{H}_B}\hat{x}_0
\me^{-\mi t\hat{H}_B}=\sum_kc^{xx}_{0,k}(t)\hat{x}_k+\sum_kc^{xp}_{0,k}(t)\hat{p}_k$,
the computation of $\epsilon_{\vec{m}}(x)$
is reduced to obtaining the coefficients $c^{xx}_{0,k}(t)$ and $c^{xp}_{0,k}(t)$ and expectations in $\me^{-\mi x\hat{H}_{\vec{m}}}\hat{\varrho}_0\me^{\mi x\hat{H}_{\vec{m}}}$ of observables of the form
\begin{equation}
\hat{h}^2\otimes(\me^{-\mi x\hat{H}_B^{\vec{m}}}\hat{r}^{\vec{m}}_k\me^{\mi x\hat{H}_B^{\vec{m}}}-\hat{x}^{\vec{m}}_0)(\me^{-\mi x\hat{H}_B^{\vec{m}}}\hat{s}^{\vec{m}}_l\me^{\mi x\hat{H}_B^{\vec{m}}}-\hat{x}_0^{\vec{m}})
\end{equation}
and
\begin{equation}
\hat{h}^2\otimes\id_{\vec{m}}\hat{r}_k(\id-\id_{\vec{m}})\hat{s}_l\id_{\vec{m}}=\delta_{k,l}\hat{h}^2\otimes\id_{\vec{m}}\hat{r}_k(\id-\id_{\vec{m}})\hat{s}_k\id_{\vec{m}}
\end{equation}
for $r,s\in\{x,p\}$.

\section{Further numerical examples of Fock space truncation}
\begin{figure*}
\centering 
\includegraphics[width=18cm]{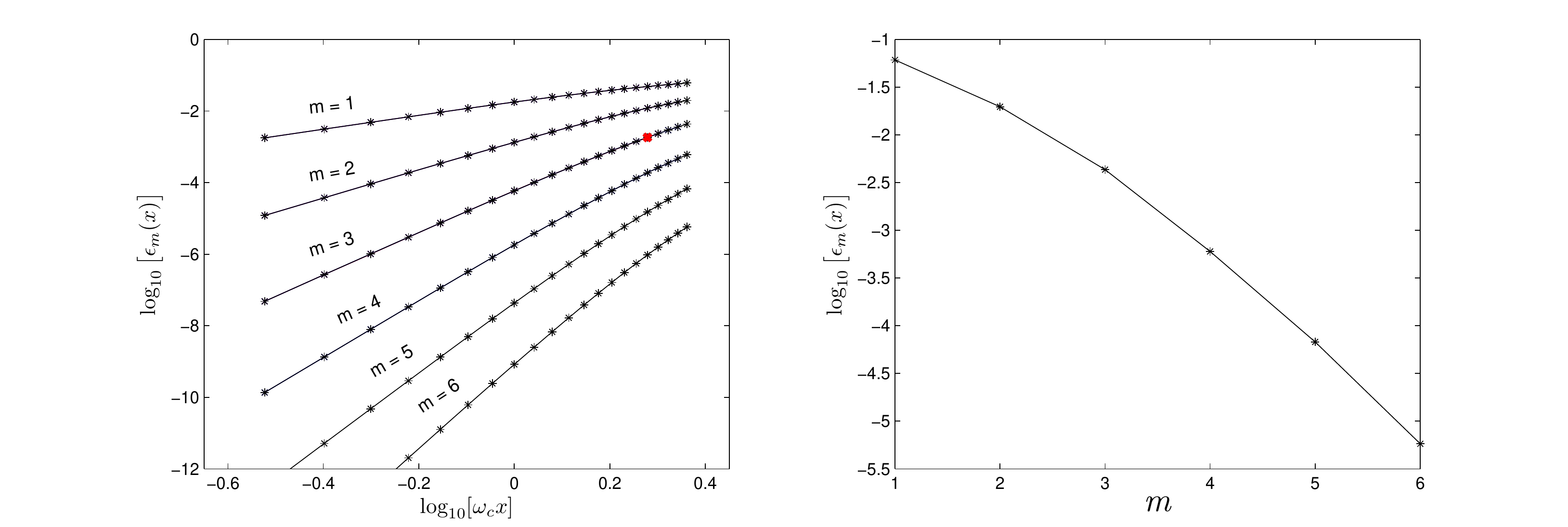} 
\caption{Fock space truncation error (Eq. (\ref{ep x rho})) for the model in Eq.\ (\ref{SB spec}) using the particle mapping with $\Delta/\omega_{c}=1$, $\alpha=0.8$, $s=1$ for initial state $\hat{\varrho}_0=\hat{\varrho}_S^0\otimes\hat{\varrho}_B^0$, $\hat{\varrho}_S^0=|\uparrow\rangle\langle\uparrow|$ and $\hat{\varrho}_B^0$ the vacuum. We truncate each local Hilbert space at the same value $m_i=m$ and $L$ has the values $3$ to $6$, but are indistinguishable (e.g. the difference between the $L=6$ and $L=3$ curve at the point denoted by a square is $4.86 \times 10^{-8}$). Lines are guides to the eye.\label{fig Particle s1}}
\end{figure*}

%
In this section, we give the chain coefficients used in the numerical simulations for the Fock space truncation and analyse further the numerical results. We start with the particle mapping of the spin-boson model:
From Eq. (\ref{PM X P Jacobi rel}), we have the relation between the $X$ and $P$ matrices and the Jacobi matrix. The coefficients of the Jacobi matrix for the spin-boson spectral density of Eq. (\ref{SB spec}) can be found in \cite{alex10} or \cite{misc14}. From Eqs. (248), (249) in \cite{misc14}, we have for $n\in \nn^0$,
\begin{align}
X_{n+1,n+1}&=\frac{\omega_c}{2}\left( 1+\frac{s^2}{(s+2 n)(2+s+2 n)} \right),\\
X_{n,n+1}=&X_{n+1,n}\\
=&\frac{\omega_c(1+n)(1+s+n)}{(s+2+2n)(3+s+2 n)}\sqrt{\frac{3+s+2 n}{1+s+2 n}},
\end{align}
and all other matrix elements zero. $\beta_0(0)$ can be found in Eq. (250) in \cite{misc14}, thus from Eq. (\ref{PM X P Jacobi rel}), we have
\be
\hat h= \omega_c\sqrt{\frac{2\alpha}{s+1}}\hat A_S.
\ee
We can now do the same for the phonon mapping written in terms of $\underline{\hat x}_n$, and $\underline{\hat p}_n$. Using Eq. (\ref{ph map dens}), we have the relation between the $X$ and $P$ matrices and the Jacobi matrix. From \cite{misc14}, we obtain the coefficients of the Jacobi matrix. Thus, using Eqs. (255), (256) in \cite{misc14}, we have for $n\in \nn^0$,
\begin{align}
X_{n+1,n+1}&=\frac{\omega_c}{2}\left( 1+\frac{s^2}{(s+4 n)(4+s+4 n)} \right),\\
X_{n,n+1}=&X_{n+1,n}
=\frac{\omega_c2 (1+n)(2+s+2n)}{(s+4+4n)(6+s+4 n)}\sqrt{\frac{6+s+4 n}{2+s+4 n}},
\end{align}
and all other matrix elements zero. $\beta_0(1)$ can be found in Eq. (257) in \cite{misc14}, thus from Eq. (\ref{ph map dens}), we have
\be
\hat h= \omega_c\sqrt{\frac{2\alpha}{s+2}}\hat A_S.
\ee
The results for the particle mapping with Ohmic spectral density are plotted in Fig.~\ref{fig Particle s1}
and for the phonon mapping in Fig.~\ref{fig}. In both cases, the plots suggest that the super
ohmic spectral densities have smaller truncation error. For the particle mappings, in the plots we have
probed the zero Kelvin thermal state (which corresponds to the chain vacuum state), where as for the phonon
mappings we have probed a squeezed vacuum state\footnote{We probed the vacuum state of the chain, which
corresponds to a squeezed vacuum state of the continuous bath of harmonic oscillators. This will be shown
in an upcoming article \cite{in_prep}.}, which is highly populated. We see that the error has slightly worse
decay with increasing $\vec m$ than in the particle mappings cases. This is intuitively what one would expect,
since more of the bath population is being truncated.

\end{document}